\documentclass[12pt]{article}
\usepackage{verbatim}
\usepackage{amsfonts}
\usepackage{graphics}
\usepackage{amsmath}
\usepackage{times}
\usepackage{appendix}
\usepackage{color}
\usepackage{soul}
\usepackage{mathtools}
\usepackage{enumerate}
\usepackage{fancyhdr,latexsym,amsmath,amsfonts,amssymb,amsbsy,amsthm,url}
\usepackage[margin=0.5in,footskip=0.25in]{geometry}
\usepackage{graphics,graphicx,epsfig}
\usepackage{breqn}
\usepackage{caption,subcaption,float,subfloat}
\usepackage{pdflscape}
\usepackage{hyperref}
\usepackage[ruled,vlined]{algorithm2e}

\usepackage[english]{babel}
\usepackage[dvipsnames]{xcolor}
\usepackage{longtable}
\usepackage{csvsimple}

\makeatletter

\renewcommand{\section}{
	\@startsection
	{section}% name
	{1}% level
	{0pt}% indent
	{1.1\baselineskip}% beforeskip
	{0.2\baselineskip}% afterskip
	{\sc \centering}% style
}

\renewcommand{\subsection}{
	\@startsection
	{subsection}% name
	{1}% level
	{0pt}% indent
	{1.1\baselineskip}% beforeskip
	{0.2\baselineskip}% afterskip
	{\sc \centering}% style
}

\renewcommand{\subsubsection}{
	\@startsection
	{subsubsection}% name
	{1}% level
	{0pt}% indent
	{1.1\baselineskip}% beforeskip
	{0.2\baselineskip}% afterskip
	{\sc \centering}% style
}

\makeatother

\usepackage[flushleft]{threeparttable}
\usepackage{rotating,booktabs,multirow}
\usepackage{colortbl}
\usepackage{makecell,cellspace,caption}

\begin{document}
	
\title{\large\sc Leverage Ratio: An empirical study of the European banking system}
\normalsize
\author{\sc{Jatin Dhingra} \thanks{Department of Mathematics, Indian Institute of Technology Guwahati, Guwahati-781039, India, e-mail: dhingra18@iitg.ac.in}
\and 
\sc{Kartikeya Singh} \thanks{Department of Mathematics, Indian Institute of Technology Guwahati, Guwahati-781039, India, e-mail: kartikey18c@iitg.ac.in}
\and 
\sc{Siddhartha P. Chakrabarty} \thanks{Department of Mathematics, Indian Institute of Technology Guwahati, Guwahati-781039, India, e-mail: pratim@iitg.ac.in, Phone: +91-361-2582606}
}

\date{}
\maketitle
\begin{abstract}

This paper empirically analyzes a dataset published by the European Banking Authority. Our main aim was to study how the Leverage Ratio is affected by adverse financial scenarios. This was be followed by observing how Leverage Ratio exposures are correlated to various other financial variables and how various regression techniques can be used to explain the correlation.


\end{abstract}

\section{Introduction}
\label{sec1}

The 2008 financial crisis was exacerbated to a great extent by the acquisition of huge leverage (by the large investment banks), while still maintaining a robust risk-based capital requirements. Accordingly, in order to avoid the recurrence of this trigger (of the crisis), a non-risk based regulatory requirement in the form of the Leverage Ratio was incorporated, in addition to the risk based regulatory requirements, in the Basel-III framework \cite{basel2014basel}. Restrictions imposed in terms of the Leverage Ratio was put in place with the intent of inhibiting a built-up of over-leveraging within the banking setup. Specifically, the Basel-III regulations defined the Leverage Ratio (expressed in terms of percentage) as the ratio of ``Capital Measure'' to ``Exposure Measure'', with the regulatory requirement being that the Leverage Ratio cannot exceed $3\%$. It may be noted that the numerator is the risk-based Tier 1 capital requirement, and the denominator is the sum-total of various defined classes of the banks' exposure (including off-balance-sheet items). 

We begin with the question of whether the Leverage Ratio increases the stability of banks. A theoretical and empirical analysis of stability of banks in the context of Leverage Ratio is examined in \cite{acosta2020leverage}. A key observation from the micro-economic model was that the imposition of the Leverage Ratio requirement, while increasing the stability of a bank, can also act as an incentive for banks to adopt more risk taking practices, and this was observed to concur with the empirical analysis on a fairly large sample of banks in the European Union (EU), for the period of 2005-2014. The model described in \cite{kiema2014does} showed that in case of banks with more conservative (in terms of risk) lending practices, the Leverage Ratio requirement may act as an inducement for such banks to diversify into risky loan portfolios, while maintaining the Leverage Ratio requirement. This in turn could contribute to the overall instability of the banking sector. In conclusion, the authors recommended a much higher level of Leverage Ratio requirement, as compared to the existing levels. Blum \cite{blum2008basel}, in his article, noted that limited supervisory authority (in terms of identification and punitive action) incentivizes under-reporting of risks by banks, and showed that Leverage Ratio would be more effective in eliciting truthful reporting of risks by the banks. An empirical examination of the impact of capital on the survival and market presence of banks, during a financial crisis is studied in \cite{berger2013does}. The two key observations were that in case of smaller banks, capital helps in terms of survival, as well as market share (in all market conditions), and that capital helps medium and large banks in terms of survival and performance, even in a crisis situation.

Another aspect that merits examination is how the risk changes as a consequence of enhancement of Leverage Ratio, with several articles in literature asserting that risk taking practices can change as a result of increase in capital requirements. An empirical study concluded that a positive relation exists between changes in capital and risk \cite{shrieves1992relationship}. Given this simultaneous behavior, most banks increase their risk exposure, concurrently with increase in capital requirement, and this holds even in case of banks which are over-capitalized. A different conclusion, was, however arrived at in a study for Swiss banks \cite{rime2001capital}, wherein it was observed that while regulatory enforcement compels banks to (naturally) increase their capital holdings, but it does not result in concurrent increase in the level of risk. In case of a study \cite{aggarwal2001impact} on the impact of the FDIC Improvement Act. (FDICIA), which required early intervention and prompt corrections, on capital requirements and credit risk, the authors concluded that increase in the capital ratios did not result in a concurrent increase in the risk. The relation between capital buffer and adjustment of portfolio risk is examined in \cite{jokipii2011bank}, with the conclusion being that this relation is positive. Further, this risk adjustment of the portfolio was observed to be contingent on the extent of capitalization.

In this paper, we empirically analyze the dataset of the EU wide stress test conducted by European Banking Authority (EBA) in 2021, for 50 banks across 15 European Economic Area (EEA), to observe the dynamics of Leverage Ratio and Leverage Ratio exposures. The remainder of this paper is organized as follows. Section \ref{sec2} presents a brief overview of the stress test dataset conducted by EBA. In Section \ref{sec3}, we empirically analyze the results of EU wide stress test to observe variations of Leverage Ratio for various banks across different European countries, during the period of 2020-23. Section \ref{sec4} empirically analyzes the results of EU wide stress test, in order to observe co-relation of Leverage Ratio exposures with other financial variables, namely, Risk exposure for Credit, Market and Operational Risk, Profit or Loss for the year, Tier 1 Capital and Net Interest Income. Finally Section \ref{sec5} presents the concluding remarks.

\section{Dataset}
\label{sec2}

The EBA conducted the stress test for 50 banks, pervading 15 EEA countries, encompassing $70\%$ of the assets held by the European banks \cite{eba2021}. The primary goal of the entire exercise was to test the resilience of the EU banking system, under two scenarios, namely, the baseline and the adverse, over a time window of three years, with a particular emphasis on taking into consideration the impact of the pandemic. The stress test conducted, identified December 2020 as the starting point of the three year time window. The first of the two cases considered in the test, namely the baseline scenario, emanating from the projections of the central banks, envisaged that the GDP of the EU will experience increases of $3.9\%$, $4.2\%$ and $2.3\%$ for 2021, 2022 and 2023 respectively. For the latter case, namely, the adverse scenario (whose design was a collaborative exercise between the European Systemic Risk Board (ESRB) and the European Central Bank (ECB)) was driven primarily by the anticipation of a long-drawn pandemic, and assumed a real decrease in the GDP of  $1.5\%$, $-1.9\%$ and $–0.2\%$ for 2021, 2022 and 2023 respectively. The dataset published by the EBA comprises of the values of several financial variables, for the time window that is being considered, and was stored in two CSV files, including several considered variables for each bank. In addition, a data dictionary file is also available, and which can be used for the purpose of deciphering the information included in the datasets.

\section{Empirical Analysis: Leverage Ratio in Baseline and Adverse Scenario}
\label{sec3}

\begin{figure}[h]
\centering 
\includegraphics[width=16cm,height=35cm,keepaspectratio]{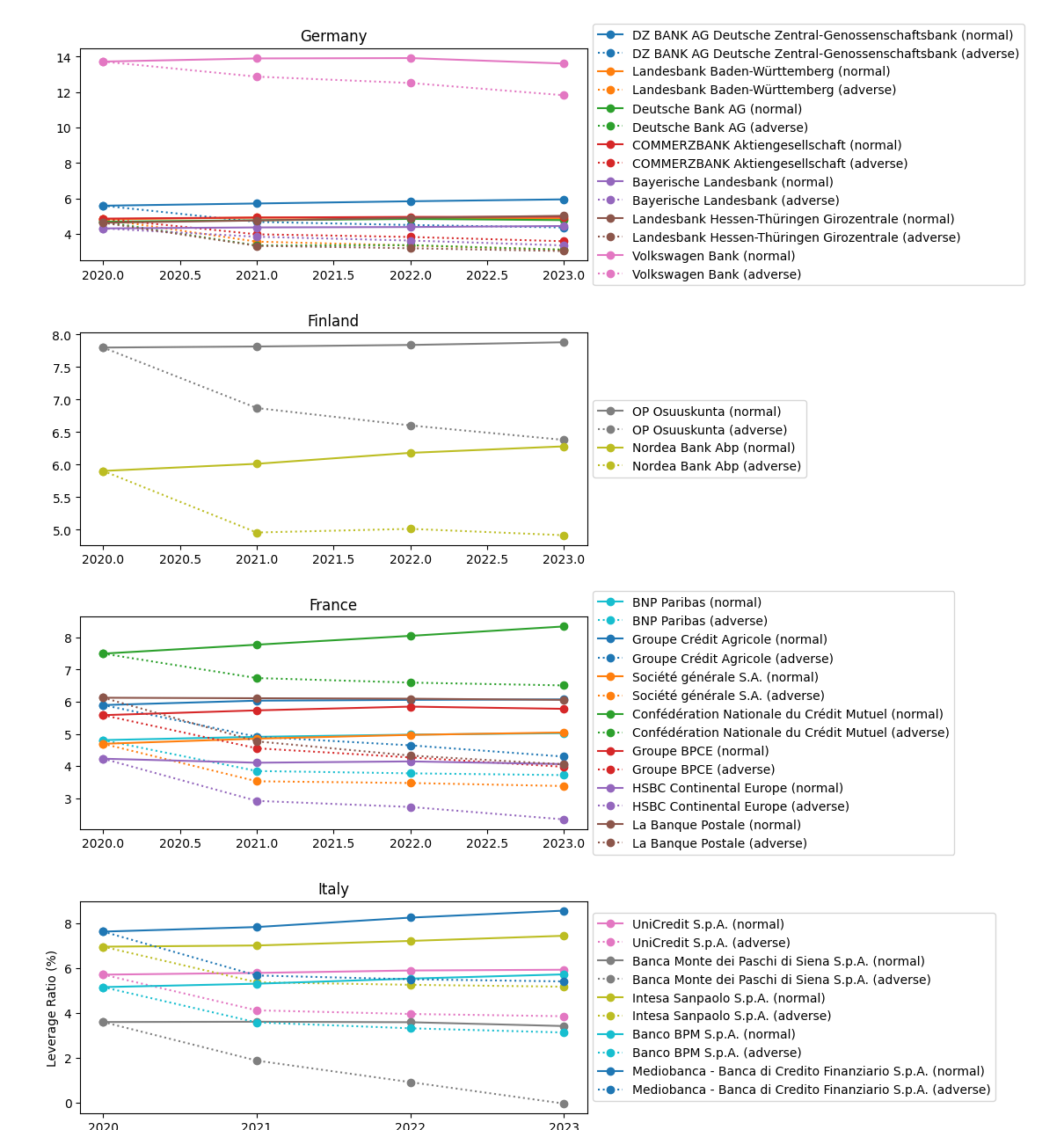}
\caption{\label{LeverageRatio} Variations of Leverage Ratio for both the scenarios}
\end{figure}

In this Section, we empirically analyze the results of the 2021 EU wide stress test, in order to observe the dynamics and variations of Leverage Ratio in case of both the baseline as well as the adverse scenario. We graphically observe the Leverage Ratio variations for various banks with each of the several European countries, for the dataset considered, during the period of 2020 to 2023. Figure \ref{LeverageRatio} shows the Leverage Ratio variations for several banks across four different countries, namely, Germany, Finland, France and Italy. It is clear from Figure \ref{LeverageRatio} that the Leverage Ratios maintained by the banks in an adverse scenario is significantly lower than the Leverage Ratio requirement in a baseline scenario, indicating that banks are more susceptible to failure in the case of an adverse scenario, as compared to the baseline scenario.  Further, the Leverage Ratio also falls below the BCBS recommendation of $3\%$, in case of few banks, in the adverse scenario. This can be attributed to the fact that the banks would hold lower capital in an adverse scenario. 

\section{Empirical Analysis: Correlation of Leverage Ratio Exposures with other Financial Variables}
\label{sec4}

In this Section, we will empirically analyze the results of the 2021 EU wide stress test, in order to observe how the Leverage Ratio exposures vary with various other financial variables, namely, Risk exposure for Credit, Market and Operational Risk, Profit or Loss for the year, Tier 1 Capital and Net Interest Income. The financial variables correspond to Codes $213501$, $213504$, $213505$, $213104$, $213110$ and $213601$,  respectively, of the Item Codes of the data dictionary file. 

The Leverage ratio exposure of a bank is measured as the sum-total of four kinds of exposures, namely, on-balance-sheet exposures, derivative exposures, securities financing transactions (SFT) exposure and off-balance sheet items \cite{basel2014basel}. This approach typically does not allow for consideration of collaterals, guarantees or other credit-risk mitigation measures adopted, to achieve any sort of reduction in the exposure measure. The on-balance sheet exposure includes all assets on the balance sheet, except derivatives and SFT assets on balance sheet, which, as already noted are considered separately. The former, namely exposure due to derivatives leads to exposures from the underlying of the derivative and exposure from Counterparty Credit Risk, while the latter, namely, the SFTs are an inseparable component of the leverage exposure, particularly securatized borrowing and lending through SFTs. Finally, the off-balance sheet items, in case of Leverage ratio were adopted from its definition incorporated in Basel II, and includes commitments, direct credit substitutes, acceptances and letter of credit (both standby and credit). Accordingly, in the following subsections, we present the results of the correlation between Leverage Ratio exposures and various other financial variables, as already identified in the preceding paragraph. The key metric for the analysis, is the coefficient of determination, $R^{2}$, which measures how much variability can be explained using the regression. 

\subsection{Risk Exposure for Credit Risk}
\begin{figure}[h]
\centering	
\includegraphics[width=13cm,height=30cm,keepaspectratio]{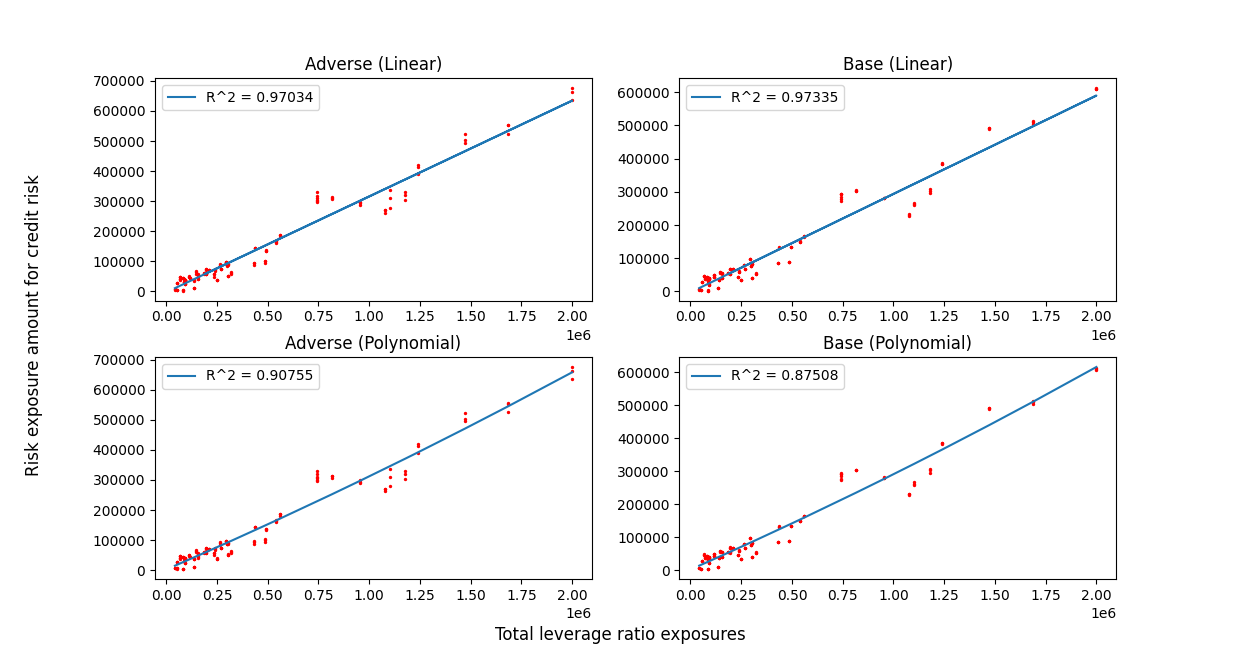}
\caption{\label{CreditRisk} Correlation with risk exposure amount for credit risk}
\end{figure}

Credit risk is defined as the likelihood of losses faced by a creditor, in the event of failure to pay back the promised amount, by the debtor. These include scenarios where the borrower fails to repay the loan amount to the bank or a bond issuer fails to make coupon payments to the bond holder. Figure \ref{CreditRisk} shows how the Leverage Ratio exposures vary with risk exposure amount for credit risk. It is clear from the figure that both linear and polynomial regression can be used to fit dataset as both result in a high value of $R^{2}$. It may be noted that risk exposure amount for credit risk increases with an increase in Leverage Ratio exposures.

\subsection{Risk Exposure for Market Risk}
\begin{figure}[h]
\centering
\includegraphics[width=13cm,height=30cm,keepaspectratio]{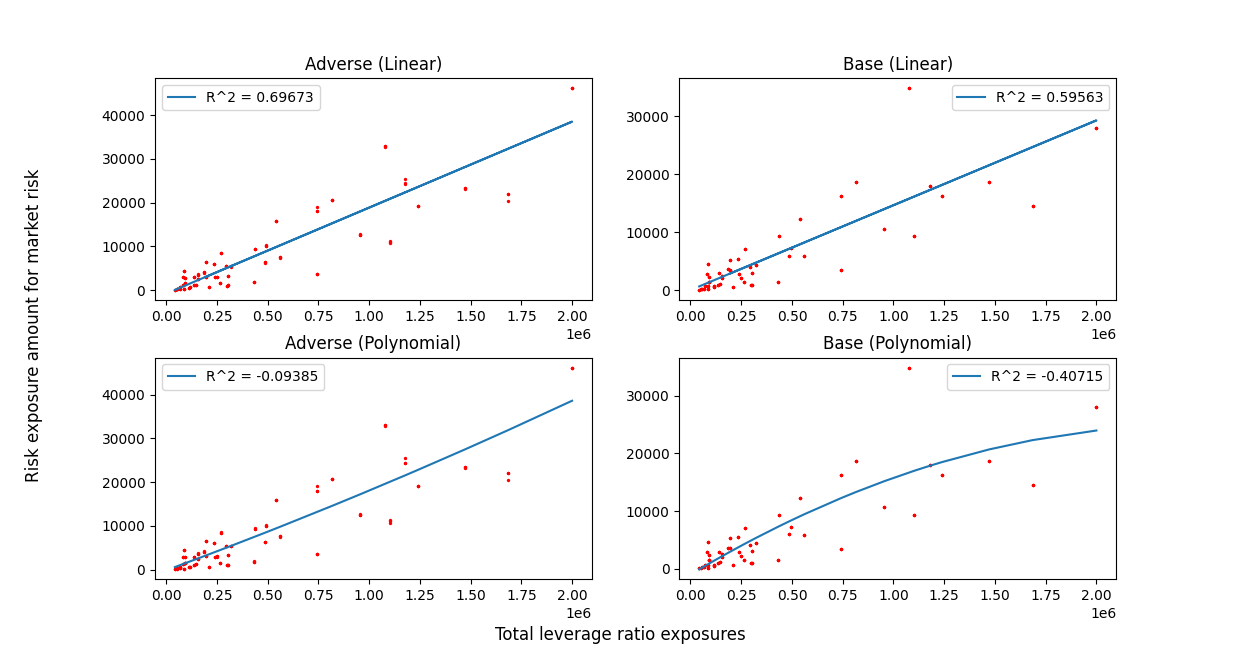}
\caption{\label{MarketRisk} Correlation with risk exposure amount for market risk}
\end{figure}

Market risk is the risk emanating from the factors that has repercussions across all kinds of investments across the market. Figure \ref{MarketRisk} shows how the Leverage Ratio exposures vary with risk exposure amount for market risk. It is clear from the figure that linear regression produces a high value of $R^{2}$, which implies that the variation between the variables can be explained by the regression. So, it is better to use linear regression to fit the dataset.

\subsection{Risk Exposure for Operational Risk}
\begin{figure}[h]
\centering
\includegraphics[width=13cm,height=30cm,keepaspectratio]{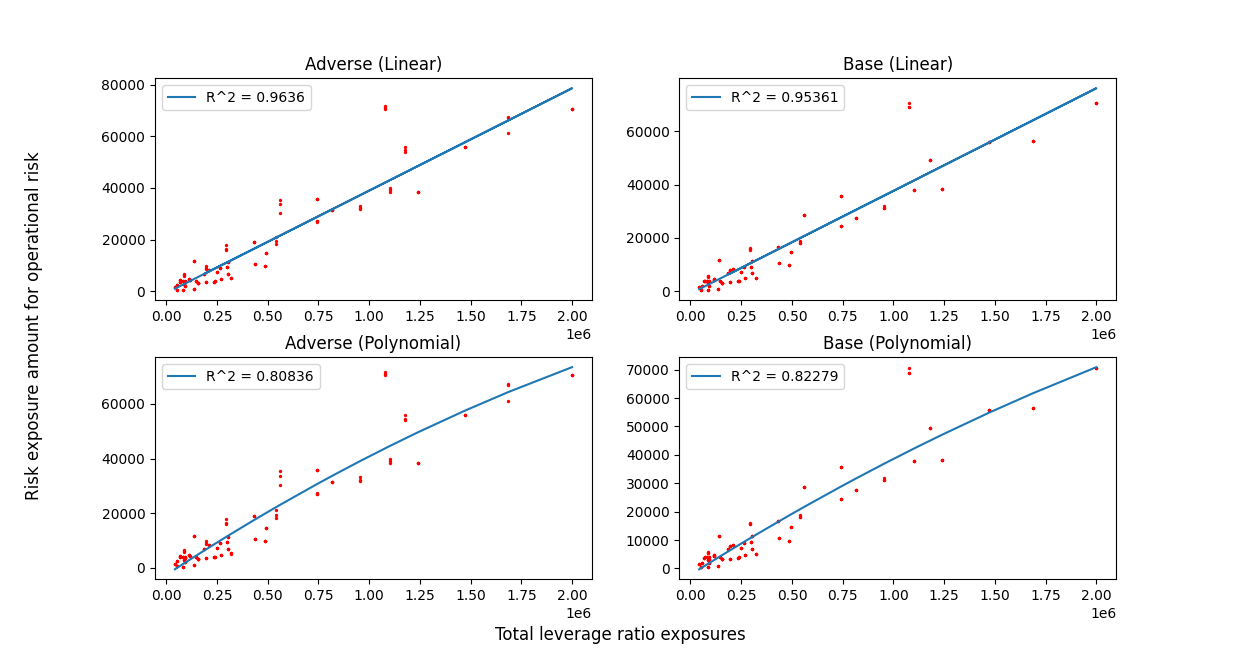}
\caption{Correlation with risk exposure amount for operational risk}
\label{OperationalRisk}
\end{figure}

Finally, operational risk refers to the risk wherein the firm would loose money, typically resulting from failure of internal processes. Figure \ref{OperationalRisk} shows how the Leverage Ratio exposures vary with risk exposure amount for Operational risk. It is clear from the figure that both linear and polynomial regression can be used to fit dataset as both produce a high value of $R^{2}$.

\subsection{Profit or Loss for the Year}
\begin{figure}[h]
\centering 
\includegraphics[width=13cm,height=30cm,keepaspectratio]{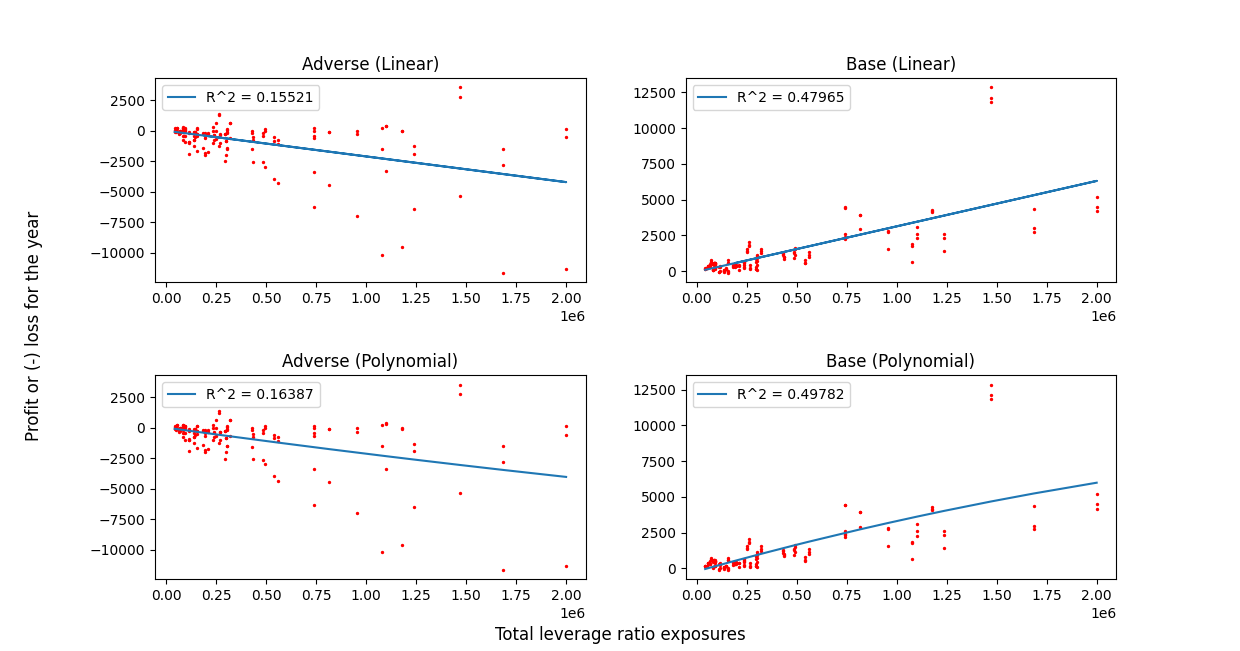}
\caption{\label{ProfitLoss} Correlation with risk exposure amount profit for  profit or loss for the year}
\end{figure}

Profit and Loss statement is an accounting metric, reflecting the revenues, as well as costs/expenses over a time window (yearly in our case). Naturally this is indicative of the firm's ability to enhance revenues and/or reducing costs/expenses. Specifically, Profit and Loss account quantifies the overall health of the firm in terms of reported ``net-profit'' earned or ``net-loss'' incurred, as the case might be. Figure \ref{ProfitLoss} shows how the Leverage Ratio exposures vary with profit or loss for the year. In the baseline scenario, profit or loss for the year increase with increase in Leverage Ratio exposure, whereas we see inverse relation in case of the adverse scenario. The value of $R^{2}$ is high for baseline scenario and small for adverse scenario, which means regression technique can be used to fit dataset in only baseline scenario. 

\subsection{Tier 1 Capital}
\begin{figure}[h]
\centering 
\includegraphics[width=13cm,height=30cm,keepaspectratio]{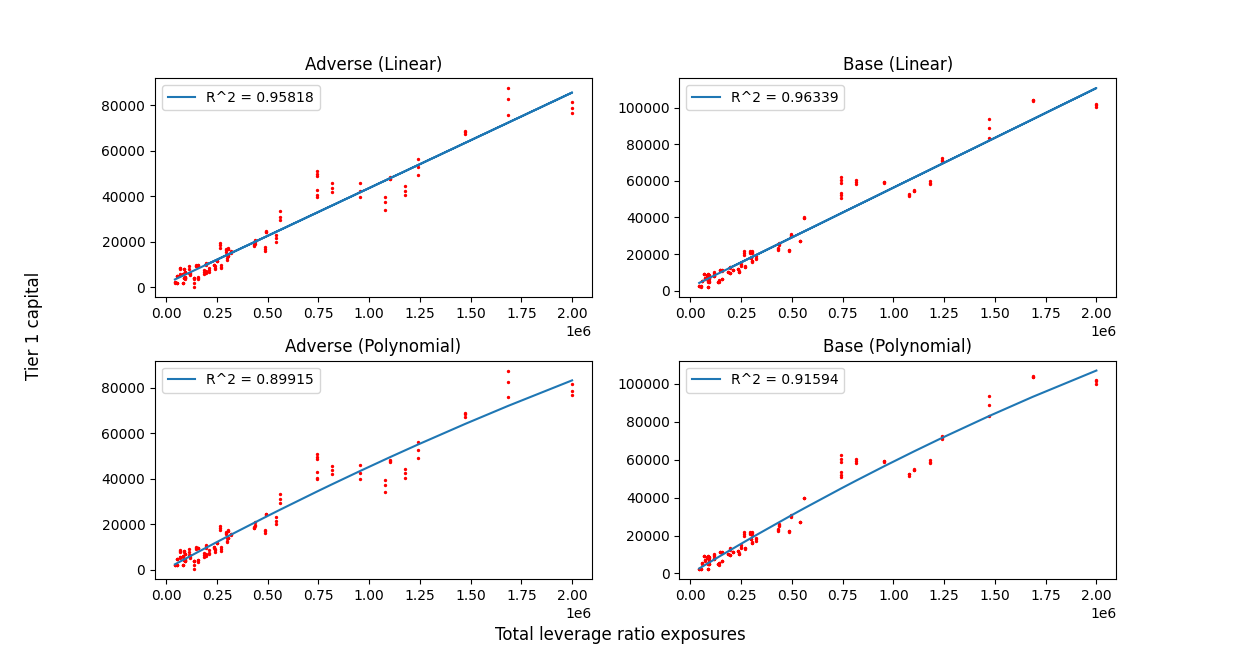}
\caption{\label{TierOne} Correlation with risk exposure amount for Tier 1 capital}
\end{figure}

Tier 1 Capital represents the pivotal regulatory capital reserve that a bank must hold, and includes (among other things) sum-total of common shares and stock surplus and retained earnings. It's strong relevance in the context of Leverage Ration stems from the fact that it appears in the numerator of its definition itself. Figure \ref{TierOne} shows how the Leverage Ratio exposures vary with Tier 1 Capital. In both scenarios, Tier 1 Capital increase with increase in Leverage Ratio exposures. Both regression techniques can be used to fit the dataset, as both have good values of $R^2$.

\subsection{Net Interest Income}
\begin{figure}[h]
\centering 
\includegraphics[width=13cm,height=35cm,keepaspectratio]{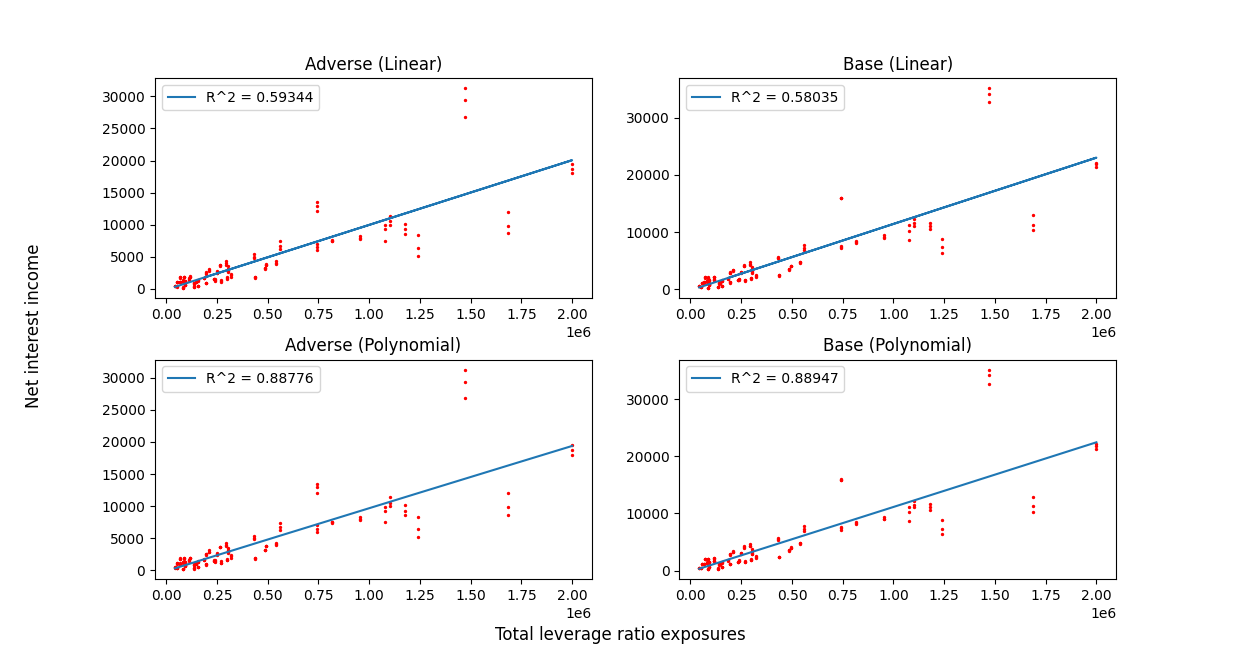}
\caption{\label{NIT} Correlation with  risk exposure amount for net interest income}
\end{figure}

Net Interest Income, as the name suggests is the difference between the revenue from assets which generate interest and the payout on liabilities emanating from interest driven assets. Examples of the former includes outstanding personal and commercial loans, while the later primarily includes interest paid on bonds and deposits. It may be noted that having a floating interest rate, makes the bank's Net Interest Income less susceptible to interest rate movements, as compared to fixed interest rates. Figure \ref{NIT} shows how the Leverage Ratio exposures vary with Net Interest Income. In both the scenarios, Net Interest Income increase with increase in Leverage Ratio exposures. Polynomial regression should be used to fit the dataset, as it produces large value of $R^{2}$ as compared to linear regression.

\subsection{$R^{2}$ Between Leverage Ratio Exposure and Other Variables}
In this Subsection, we analyze how the Leverage Ratio exposures is correlated to various other financial variables considered, using different regression techniques, namely, linear, polynomial, decision tree, SVM and random forest regression. Tables \ref{tab:base} and \ref{tab:adv} represents the $R^{2}$ (coefficient of determination) values in the baseline and the adverse scenario of the regression between Leverage Ratio exposures and financial variables discussed in previous sections, using different regression techniques. The value of $R^{2}$ is a measure of the effectiveness of the regression technique, and values close to $1$ reflects that the regression is able to predict the values with high accuracy.
The variable code in the Tables \ref{tab:base} and \ref{tab:adv} correspond to the variable codes as found in \cite{eba2021}. The values for all the variables have been reported in the Appendix \ref{secA1}.
{\small
	\begin{filecontents*}{r2_base2.csv}
Variable Code,Linear,Polynomial,SVM,Random Forest,Decision Tree
213104,0.47965,0.49782,0.45151,0.76388,0.96019
213110,0.96339,0.91594,0.97962,0.99271,0.99614
213501,0.97335,0.87508,0.97261,0.9974,0.99798
213504,0.59563,-0.40715,0.70123,0.93424,0.97239
213505,0.95361,0.82279,0.96662,0.9823,0.98527
213601,0.58035,0.88947,0.5796,0.80584,0.96465
\end{filecontents*}
\csvreader[
	longtable = |p{0.14\linewidth}|p{0.14\linewidth}|p{0.14\linewidth}|p{0.14\linewidth}|p{0.14\linewidth}|p{0.14\linewidth}|,
	table head=\caption{$R^2$ values in baseline scenario.\label{tab:base}}\\ \hline \bfseries{Variable Code} & \bfseries{Linear} & \bfseries{Polynomial} & \bfseries{SVM} & \bfseries{Random Forest}  & \bfseries{Decision Tree} \\\hline,
	late after last line=\\\hline,
	]{r2_base2.csv}{}{\csvlinetotablerow}
}
{\small
	\begin{filecontents*}{r2_adv2.csv}
Variable Code,Linear,Polynomial,SVM,Random Forest,Decision Tree
213104,0.15521,0.16387,-0.05407,-0.33368,-1.34394
213110,0.95818,0.89915,0.97206,0.99059,0.9834
213501,0.97034,0.90755,0.98179,0.99494,0.99488
213504,0.69673,-0.09385,0.71964,0.94434,0.96403
213505,0.9636,0.80836,0.96447,0.97326,0.97713
213601,0.59344,0.88776,0.5868,0.75492,0.92874
\end{filecontents*}
\csvreader[
	longtable = |p{0.14\linewidth}|p{0.14\linewidth}|p{0.14\linewidth}|p{0.14\linewidth}|p{0.14\linewidth}|p{0.14\linewidth}|,
	table head=\caption{$R^2$ values in adverse scenario.\label{tab:adv}}\\ \hline \bfseries{Variable Code} & \bfseries{Linear} & \bfseries{Polynomial} & \bfseries{SVM} & \bfseries{Random Forest}  & \bfseries{Decision Tree} \\\hline,
	late after last line=\\\hline,
	]{r2_adv2.csv}{}{\csvlinetotablerow}
}

\section{Conclusion}
\label{sec5}

In this paper, we introduced the notion of Leverage Ratio as described in Basel-III, and discussed its importance in the context of stability of the banking system, as well as its repercussion on the risk-taking tendencies in the banking sector. Then we empirically analyzed the EU-wide stress test dataset to observe how Leverage Ratio varies in a baseline and an adverse scenario. We observed that Leverage Ratio maintained by banks is significantly lower in case of an adverse scenario, as compared to a normal scenario, implying that bank stability increases with an increase in Leverage Ratio. In the preceding section we studied the correlation of the Leverage Ratio with several identified financial variables, and used various regression techniques to find a correlation between Leverage Ratio and all other variables available in the EU-wide stress test dataset.  

\begin{appendices}
	
\section{Tables}
\label{secA1}

{\tiny
	\begin{filecontents*}{r2_base1.csv}
Variable Code,Linear,Polynomial,SVM,Random Forest,Decision Tree
213101,0.58035,0.88947,0.5796,0.80584,0.96465
213102,0.67944,-2.1439,0.75229,0.99841,0.99923
213103,0.74837,0.4389,0.68999,0.68626,0.74638
213104,0.47965,0.49782,0.45151,0.76388,0.96019
213105,-0.01997,-0.02685,0.02907,0.50825,0.53015
213106,0.94343,0.89512,0.96389,0.99184,0.99678
213107,0.98232,0.9003,0.98104,0.99569,0.99637
213108,0.00365,-0.10463,-0.05466,0.92168,0.94657
213109,-0.00053,-0.10225,-0.07666,0.91544,0.9411
213110,0.96339,0.91594,0.97962,0.99271,0.99614
213112,-0.13294,0.06468,0.01288,0.91474,0.92368
213113,-0.18112,0.06026,-0.04759,0.89465,0.90461
213114,1.0,1.0,1.0,1.0,1.0
213115,0.2731,0.29615,0.13554,0.91363,1.0
213116,1.0,1.0,1.0,1.0,1.0
213501,0.97335,0.87508,0.97261,0.9974,0.99798
213502,0.79462,-0.48142,0.77047,0.98168,0.99049
213503,0.97187,0.87784,0.97205,0.99742,0.99798
213504,0.59563,-0.40715,0.70123,0.93424,0.97239
213505,0.95361,0.82279,0.96662,0.9823,0.98527
213506,-0.1447,0.03387,0.07451,0.90905,0.97481
213507,0.98238,0.90022,0.98092,0.99569,0.99637
213508,0.98232,0.9003,0.98104,0.99569,0.99637
213509,0.98255,0.90107,0.98073,0.99521,0.99589
213601,0.58035,0.88947,0.5796,0.80584,0.96465
213602,0.27551,0.05251,0.49163,0.7683,0.93573
213603,-0.011,-2.09151,0.38872,0.72288,0.87308
213604,0.29316,0.31783,0.2352,0.79947,0.93322
213605,0.90517,0.38039,0.91685,0.98722,0.99715
213606,0.67944,-2.1439,0.75229,0.99841,0.99923
213608,0.28947,0.35384,0.38443,0.96113,0.98983
213609,0.87595,0.81023,0.86496,0.94855,0.98927
213610,0.74837,0.4389,0.68999,0.68626,0.74638
213611,0.9717,0.82507,0.96199,0.99744,0.99809
213612,0.45174,0.45035,0.43082,0.75378,0.96065
213613,0.38466,0.45497,0.38825,0.73174,0.96192
213615,0.47965,0.49782,0.45151,0.76388,0.96019
213616,0.43719,0.09697,0.3839,0.82575,0.988
213617,0.47467,0.50115,0.51687,0.70127,0.90244
213618,-0.00834,-0.01709,-0.14015,0.83582,1.0
213619,0.0,0.0,0.0,0.0,1.0
213701,0.9668,0.9104,0.97988,0.99275,0.99639
213702,0.94343,0.89512,0.96389,0.99184,0.99678
213703,0.6535,0.25496,0.62323,0.91114,0.98548
213704,-0.02514,0.02517,0.00157,0.1545,1.0
213705,-0.10831,0.51742,-0.11605,0.76909,0.96049
213706,0.10808,-0.39635,0.05889,0.78878,0.99186
213707,0.12442,-1.11785,0.06438,0.77719,0.99991
213708,0.08865,0.10345,0.01053,0.85338,0.95499
213709,-0.20911,-0.29699,0.09491,0.55217,0.80107
213710,0.15507,-1.33775,-0.11888,0.82366,0.9947
213711,0.0,-0.01526,0.0,1.0,1.0
213712,0.37357,-0.38006,0.5356,0.36942,0.41234
213713,0.7546,0.19601,0.8734,0.87745,0.94247
213714,0.85131,0.20061,0.87397,0.95161,0.98052
213715,0.46029,0.09859,0.75635,0.76376,0.99831
213716,-0.08565,0.05913,0.04725,-0.26915,-0.25075
213717,0.65732,-0.55359,0.81094,0.99033,0.99954
213718,-0.17466,-0.14739,-0.04375,0.90175,0.96595
213719,0.17764,-0.27446,0.17898,0.67303,0.75924
213720,0.56706,0.29844,0.15772,-1.30519,-1.05013
213721,-0.41517,0.27753,-0.00991,0.83096,1.0
213722,-0.00865,0.0,-0.01247,0.9069,1.0
213723,1.0,1.0,1.0,1.0,1.0
213724,-0.00675,0.02519,-0.18723,0.77907,0.97244
213725,-0.11911,0.11795,-0.27791,0.81064,0.9623
213726,1.0,1.0,1.0,1.0,1.0
213727,-0.07636,-136.5441,-0.00663,0.8887,0.99603
213728,0.02495,0.04059,-0.02109,0.99871,0.99957
213729,0.03724,-0.00902,-0.04491,0.89987,0.92182
213730,-9.08599,-0.40571,-1.59121,0.47412,1.0
213731,-2.12758,-0.07816,-0.36345,0.87539,0.94844
213732,0.50366,0.12315,0.44023,0.95604,0.95255
213733,0.36728,0.10949,0.46721,0.96427,1.0
213734,-1.99561,-0.05667,-0.05305,0.84452,1.0
213735,0.22075,0.18097,0.43441,0.95222,1.0
213790,0.27356,0.17746,-0.07216,0.59423,0.63511
213791,0.15691,0.17089,-0.05319,0.5698,0.60518
213736,-0.00452,-0.12644,-0.12029,0.69457,0.63416
213737,0.0,0.0,0.0,0.0,0.0
213738,-0.00148,-43.68276,0.00174,0.96997,0.74928
213739,0.13436,-0.89928,-0.02425,0.29406,0.13225
213740,0.13904,-0.75555,-0.01032,0.28232,0.1118
213741,-0.00638,-10913.88441,-0.03735,0.73419,0.81538
213742,0.01855,0.03234,-0.02218,0.99361,0.99634
213743,-0.0817,0.01706,0.00155,0.75065,0.99996
213744,-0.01991,-0.54359,-0.03107,0.75835,0.62585
213792,0.0,-0.03416,0.0,0.0,1.0
213745,0.43493,0.71489,0.34882,0.87704,0.92302
213746,0.32494,0.56199,0.24473,0.85098,0.91701
213747,1.0,1.0,1.0,1.0,1.0
213748,0.15418,-155.73056,0.9452,0.98692,0.9931
213749,0.47291,0.23444,0.71938,0.95046,0.95246
213750,1.0,1.0,1.0,1.0,1.0
213751,0.96339,0.91594,0.97962,0.99271,0.99614
213752,0.95672,0.79169,0.95974,0.98437,0.99331
213753,0.95971,0.76074,0.96144,0.98943,0.99827
213754,0.58254,-0.4193,0.49002,0.81196,0.84648
213755,-0.28143,0.07566,0.30503,-0.18342,-1.28108
213756,-1.10375,0.05177,-0.02947,0.09397,-0.03858
213757,-2.08343,0.10212,-0.14699,-7.27674,-27.88345
213758,0.98238,0.90022,0.98092,0.99569,0.99637
213759,-0.00678,0.02192,0.01245,0.66369,0.66605
213760,0.00683,0.06598,-0.01101,0.47436,0.51013
213761,0.00365,-0.10463,-0.05466,0.92168,0.94657
213762,0.00487,-0.09763,-0.10124,0.95539,0.98025
213763,-0.00354,-0.08646,-0.15204,0.95013,0.97813
213764,0.94164,0.89497,0.96033,0.99104,0.9961
213765,0.96235,0.91795,0.97766,0.99201,0.99543
213766,0.9661,0.91183,0.97912,0.9922,0.99587
213767,-0.00053,-0.10225,-0.07666,0.91544,0.9411
213768,0.00365,-0.09273,-0.12135,0.95141,0.97682
213769,-0.00455,-0.08043,-0.15658,0.94527,0.97393
213770,1.0,1.0,1.0,1.0,1.0
213771,0.2731,0.29615,0.13554,0.91363,1.0
213772,1.0,1.0,1.0,1.0,1.0
213773,1.0,1.0,0.99488,0.9996,1.0
213774,1.0,0.99999,0.99484,0.99961,1.0
213775,-0.13294,0.06468,0.01288,0.91474,0.92368
213776,-0.18112,0.06026,-0.04759,0.89465,0.90461
213777,-1.1890417140759336e+28,0.01748,-2.568832139658387e+27,-2.75,-44.8
213778,-0.00525,-0.00382,-0.04422,0.78186,0.8719
213779,-0.33688,0.12732,-0.16769,0.26566,0.34029
213780,0.84003,0.66858,0.98691,0.9988,1.0
213781,-0.00903,-0.01131,-0.09194,0.4525,0.47059
213782,-0.01993,-0.0139,-0.06898,0.92375,1.0
213783,-0.07337,0.06718,0.11555,0.75341,0.83035
213784,0.01408,0.0098,0.063,0.95335,0.98113
213785,-0.00472,0.03504,0.06977,0.92268,0.98704
213786,0.01408,0.0098,0.063,0.95335,0.98113
213787,-0.00472,0.03504,0.06979,0.92268,0.98704
213788,-0.121,0.09814,0.09663,0.78256,0.87089
213789,-0.0873,0.0879,0.1079,0.79108,0.8801
\end{filecontents*}
\csvreader[
	longtable = |p{0.14\linewidth}|p{0.14\linewidth}|p{0.14\linewidth}|p{0.14\linewidth}|p{0.14\linewidth}|p{0.14\linewidth}|,
	table head=\caption{$R^2$ values in baseline scenario.\label{tab:base1}}\\ \hline \bfseries{Variable Code} & \bfseries{Linear} & \bfseries{Polynomial} & \bfseries{SVM} & \bfseries{Random Forest}  & \bfseries{Decision Tree} \\\hline,
	late after last line=\\\hline,
	]{r2_base1.csv}{}{\csvlinetotablerow}
}
{\tiny
		\begin{filecontents*}{r2_adv1.csv}
Variable Code,Linear,Polynomial,SVM,Random Forest,Decision Tree
213101,0.59344,0.88776,0.5868,0.75492,0.92874
213102,-0.13668,-0.48984,-0.272,-1.65509,-2.33107
213103,0.76311,0.46647,0.57693,0.29402,0.24461
213104,0.15521,0.16387,-0.05407,-0.33368,-1.34394
213105,-0.03107,0.00521,0.07777,0.45498,0.45663
213106,0.92656,0.8533,0.94845,0.99187,0.98352
213107,0.98223,0.92977,0.98864,0.99405,0.99368
213108,0.03153,-0.06687,0.01202,0.94457,0.95981
213109,0.02876,-0.07756,0.00674,0.9515,0.97054
213110,0.95818,0.89915,0.97206,0.99059,0.9834
213112,-0.24458,0.05064,-0.00135,0.79841,0.82016
213113,-0.27121,0.0387,-0.03975,0.81792,0.84719
213114,1.0,1.0,1.0,1.0,1.0
213115,0.2731,0.29615,0.13554,0.91363,1.0
213116,1.0,1.0,1.0,1.0,1.0
213501,0.97034,0.90755,0.98179,0.99494,0.99488
213502,0.76521,-0.11381,0.73544,0.92255,0.88429
213503,0.96833,0.90567,0.98063,0.99591,0.99615
213504,0.69673,-0.09385,0.71964,0.94434,0.96403
213505,0.9636,0.80836,0.96447,0.97326,0.97713
213506,-0.15434,-0.00563,0.0674,0.92268,0.971
213507,0.98223,0.92942,0.98876,0.99428,0.99378
213508,0.98223,0.92977,0.98864,0.99405,0.99368
213509,0.98236,0.92993,0.98869,0.99399,0.99343
213601,0.59344,0.88776,0.5868,0.75492,0.92874
213602,0.23664,0.10411,0.49349,0.72831,0.90813
213603,-0.04211,-1.84858,0.3709,0.7044,0.8643
213604,0.25271,0.3225,0.19015,0.80101,0.96666
213605,0.89448,0.40927,0.90476,0.98737,0.99439
213606,-0.13668,-0.48984,-0.272,-1.65509,-2.33107
213608,0.26066,0.35849,0.39044,0.93029,0.96084
213609,0.83503,0.82655,0.834,0.95734,0.96181
213610,0.76311,0.46647,0.57693,0.29402,0.24461
213611,0.9633,0.8438,0.9297,0.95527,0.95459
213612,0.1908,0.1675,-0.01428,-0.17878,-0.89377
213613,0.2394,0.2552,-0.02726,0.04828,-0.10717
213615,0.15521,0.16387,-0.05407,-0.33368,-1.34394
213616,0.01582,-0.07618,0.11076,0.38271,0.29793
213617,0.25998,0.15174,-0.04549,-0.41917,-1.62423
213618,-0.00813,-0.01804,-0.14007,0.83584,1.0
213619,-82.23336,-0.43079,-6.87159,-57.95867,-47.83539
213701,0.96441,0.90733,0.97598,0.99217,0.98706
213702,0.92656,0.8533,0.94845,0.99187,0.98352
213703,0.6535,0.25496,0.62323,0.91114,0.98548
213704,-0.02514,0.02517,0.00157,0.1545,1.0
213705,-0.52466,0.34994,-0.61632,0.60421,0.95816
213706,0.12551,-0.27782,0.07817,0.77974,0.99924
213707,0.1724,-0.35194,0.14431,0.78786,0.99861
213708,0.10992,0.13496,0.00728,0.83968,0.96047
213709,-0.26408,-0.28276,0.19597,0.77504,0.96927
213710,0.15547,-1.27377,-0.11898,0.82353,0.99458
213711,0.0,-0.01526,0.0,1.0,1.0
213712,0.43152,-0.27921,0.58264,0.20153,0.22351
213713,0.8643,0.2026,0.97045,0.96367,0.98615
213714,0.94507,-1.00136,0.98867,0.99295,0.99647
213715,0.49859,0.11665,0.87306,0.81494,0.98994
213716,-0.23182,0.12067,0.08826,-0.26915,-0.25075
213717,0.6502,-0.78136,0.81147,0.99032,0.99954
213718,0.59159,0.28051,0.53275,0.67767,0.72315
213719,0.36707,-0.05609,0.32947,0.90912,0.99909
213720,0.55667,0.28388,0.05543,-1.29614,-1.01124
213721,-0.75813,0.24205,0.06049,0.93588,1.0
213722,-0.00865,0.0,-0.01247,0.9069,1.0
213723,1.0,1.0,1.0,1.0,1.0
213724,-0.00726,0.02732,-0.18685,0.77911,0.97248
213725,-0.11986,0.11851,-0.27687,0.81069,0.96235
213726,0.0,-0.03448,0.0,0.0,1.0
213727,-0.05722,-0.4292,-0.22105,0.75691,0.93668
213728,0.03086,0.03955,-0.02576,0.99717,0.99956
213729,-0.44648,0.13409,-0.08303,0.68894,0.94671
213730,-28.0607,-0.44716,-5.15562,-0.63044,1.0
213731,-2.29011,0.00466,-0.35022,0.87539,0.94844
213732,0.60447,0.13034,0.64569,0.96139,0.93522
213733,0.36728,0.10949,0.46721,0.96427,1.0
213734,-1.99561,-0.05667,-0.05305,0.84452,1.0
213735,0.22075,0.18097,0.43441,0.95222,1.0
213790,0.58934,0.11216,0.72207,0.96593,0.95306
213791,0.42244,0.20641,0.38765,0.59391,0.63384
213736,0.53363,-1.24723,0.53937,0.36012,0.26877
213737,0.0,0.0,0.0,0.0,0.0
213738,-16.46446,-139.21537,-7.77106,-335.78901,-954.48611
213739,0.61073,-2.15571,0.61117,0.36273,0.26903
213740,0.58967,-1.12064,0.60405,0.59453,0.47866
213741,0.1549,-3647.9413,0.05484,-0.10397,0.06623
213742,0.02751,0.03317,-0.03164,0.98383,0.9829
213743,-0.08843,0.01685,0.00102,0.74876,1.0
213744,-0.03088,-1.22321,-0.01613,0.80616,0.67941
213792,0.0,0.0,0.0,0.0,1.0
213745,0.43558,0.71511,0.34786,0.87718,0.92307
213746,0.32553,0.56256,0.24388,0.85099,0.91702
213747,1.0,1.0,1.0,1.0,1.0
213748,0.15418,-155.73056,0.9452,0.98692,0.9931
213749,0.47281,0.23367,0.71868,0.9503,0.95201
213750,1.0,1.0,1.0,1.0,1.0
213751,0.95818,0.89915,0.97206,0.99059,0.9834
213752,0.94339,0.83621,0.95679,0.9806,0.99181
213753,0.96018,0.77276,0.96433,0.98952,0.99827
213754,0.03592,0.14773,0.77537,0.78246,0.78924
213755,0.04819,-0.12474,0.03042,0.53958,0.75245
213756,-0.15767,-0.07018,-0.09333,0.59565,0.93143
213757,-2.08343,0.10212,-0.14699,-7.27674,-27.88345
213758,0.98223,0.92942,0.98876,0.99428,0.99378
213759,-0.00448,0.01144,0.0224,0.50108,0.50915
213760,0.2366,-0.49877,0.22753,0.40984,0.73517
213761,0.03153,-0.06687,0.01202,0.94457,0.95981
213762,0.03259,-0.07259,0.00674,0.9551,0.97481
213763,0.03008,-0.08752,-0.0129,0.95287,0.96667
213764,0.91321,0.84353,0.92109,0.99045,0.98396
213765,0.95126,0.88952,0.96093,0.98931,0.98387
213766,0.96104,0.89953,0.96954,0.99111,0.98774
213767,0.02876,-0.07756,0.00674,0.9515,0.97054
213768,0.03282,-0.07357,0.01117,0.96091,0.98372
213769,0.03054,-0.08589,-0.00619,0.96037,0.97788
213770,1.0,1.0,1.0,1.0,1.0
213771,0.2731,0.29615,0.13554,0.91363,1.0
213772,1.0,1.0,1.0,1.0,1.0
213773,1.0,1.0,0.99488,0.9996,1.0
213774,1.0,0.99999,0.99484,0.99961,1.0
213775,-0.24458,0.05064,-0.00135,0.79841,0.82016
213776,-0.27121,0.0387,-0.03975,0.81792,0.84719
213777,-1.1890417140759336e+28,0.01748,-2.568832139658387e+27,-2.75,-44.8
213778,-0.03883,0.00022,-0.06995,0.78186,0.8719
213779,-0.33688,0.12732,-0.16769,0.26566,0.34029
213780,0.84003,0.66858,0.98691,0.9988,1.0
213781,-0.00903,-0.01131,-0.09194,0.4525,0.47059
213782,-0.01993,-0.0139,-0.06898,0.92375,1.0
213783,-0.07896,0.06505,0.11272,0.75341,0.83035
213784,0.01415,0.00908,0.06138,0.95397,0.98127
213785,-0.00436,0.03445,0.06894,0.92329,0.9871
213786,0.01415,0.00908,0.06154,0.95397,0.98127
213787,-0.00436,0.03445,0.06891,0.92329,0.9871
213788,-0.12939,0.09921,0.08856,0.78189,0.87049
213789,-0.09341,0.08787,0.10307,0.79066,0.87986
\end{filecontents*}
\csvreader[
		longtable = |p{0.14\linewidth}|p{0.14\linewidth}|p{0.14\linewidth}|p{0.14\linewidth}|p{0.14\linewidth}|p{0.14\linewidth}|,
		table head=\caption{$R^2$ values in adverse scenario.\label{tab:adv1}}\\ \hline \bfseries{Variable Code} & \bfseries{Linear} & \bfseries{Polynomial} & \bfseries{SVM} & \bfseries{Random Forest}  & \bfseries{Decision Tree} \\\hline,
		late after last line=\\\hline,
		]{r2_adv1.csv}{}{\csvlinetotablerow}
}
	
\end{appendices}

\nocite{*}

\bibliographystyle{elsarticle-num}

\bibliography{BIBLIO}
	
\end{document}